# Copper waveguide cavities with reduced surface loss for coupling to superconducting qubits

Daniela F. Bogorin[1], D. T. McClure[2], Matthew Ware[1], B. L. T. Plourde[1]*
[1]Department of Physics, Syracuse University, Syracuse, New York, 13244-1130, USA

[2]IBM T.J. Watson Research Center, Yorktown Heights, New York, 10598, USA
*bplourde@syr.edu

*Abstract*—**Significant improvements in superconducting qubit coherence times have been achieved recently with three-dimensional microwave waveguide cavities coupled to transmon qubits. While many of the measurements in this direction have utilized superconducting aluminum cavities, other recent work has involved qubits coupled to copper cavities with coherence times approaching 0.1 ms. The copper provides a good path for thermalizing the cavity walls and qubit chip, although the substantial cavity loss makes conventional dispersive qubit measurements challenging. We are exploring various approaches for improving the quality factor of three-dimensional copper cavities, including electropolishing and coating with superconducting layers of tin. We have characterized these cavities on multiple cooldowns and found the tin-plating to be robust. In addition, we have performed coherence measurements on transmon qubits in these cavities and observed promising performance**.

## I. INTRODUCTION

Artificial atoms formed from Josephson junction-based superconducting circuits are a promising system for implementing quantum information processing [1, 2]. Over the past 15 years, the performance of these superconducting qubits has improved dramatically, with increases in coherence times by many orders of magnitude [3]. During this time there have been many key innovations, including the development of the field of circuit quantum electrodynamics (cQED), where microwave resonant cavities are coupled to superconducting qubits, resulting in a scheme for preparing, protecting, and measuring the qubit state [4, 5]. Such cavities can also be used for storing and exchanging quantum information [6, 7].

Some of the recent substantial improvements in coherence times of superconducting qubits have been achieved through the use of three-dimensional microwave waveguide cavities in place of the more conventional planar two-dimensional superconducting resonators [8]. In this approach a transmon qubit [9] is fabricated on a small substrate that is placed inside a hollow waveguide cavity near an electric field antinode for one of the resonant modes. One of the dominant loss mechanisms in superconducting planar resonant structures comes from oxides at the superconductor-substrate and substrate-air interfaces [10]. Thus, the elimination of much of the substrate in the three-dimensional cavity-qubit systems has greatly reduced this loss channel and resulted in the significant enhancements in qubit energy relaxation lifetime $T_1$. Qubit coherence times $T_2^*$ in three-dimensional systems have also been enhanced through improved thermalization of the cavity and electrical leads to reduce dephasing processes caused by stray photons occupying cavity modes that couple to the qubit [11].

While many of the recent experiments with waveguide cavity-qubit systems have utilized superconducting Al cavities, other qubit measurements have been performed with normal-metal copper cavities for ensuring a strong thermal pathway for cooling the cavity walls and qubit chip [12]. This work also resulted in long $T_1$ times and $T_2^*$ times approaching 0.1 ms. Although the surface loss in a copper cavity results in substantially lower quality factors compared to a superconducting cavity, for sufficiently large cavity-qubit detuning, the impact on the qubit $T_1$ due to the Purcell effect [13] can be minimized. However, the broad linewidth for a copper cavity relative to the typical state-dependent dispersive shift from a superconducting qubit tends to limit measurement fidelity [14]. Thus, the ability to reduce the surface loss in the cavity and increase the quality factor while preserving the good thermal conducting properties of a copper structure would be advantageous [15].

Here we report our progress with developing surface treatments for copper waveguide cavities as well as techniques for coating the copper with thin superconducting layers of Sn. We present low-temperature measurements of the quality factors of these cavities as well as coherence experiments for superconducting qubits coupled to these copper cavities.

## II. CAVITY FABRICATION

The cavity/qubit geometry has a similar design to the one used in Ref. [12]. The Cu cavity block is split into two halves along the middle allowing the qubit chip to be placed in the center (Fig. 1), thus coupling to the electric field antinode of the TE101 mode. The excitation and readout of the cavity is done with two SMA connectors, one in each cavity half, with pins extending into apertures in one of the cavity walls (Fig. 1). By varying the penetration of each of the SMA pins into the aperture, it is possible to control the coupling strength between the cavity resonance and the measurement circuitry. Our microwave cavities were machined from oxygen-free high-conductivity (OFHC) Cu. We present measurements taken on two Cu cavities with the same layout and design dimensions of $18.6 \times 15.4 \times 4.1$ mm$^3$. The bare cavity resonance with no qubit chip was 12.607 GHz for cavity-I and 12.833 GHz for cavity-II. The small differences in frequency are due to minor variations during the machining of the two cavities.

Following the cavity fabrication, we start by cleaning the Cu blocks in order to remove any debris and grease used in the machining process. For this we use an ultrasonic cleaner and immerse the blocks in acetone for a total of 10 minutes, changing the acetone halfway through, then repeat the process for 5 minutes in isopropyl alcohol. We then dry the Cu blocks with $N_2$ gas. We next use an acid etch to remove any oxide layers on the Cu surface and as a first step in reducing the surface roughness. We use a standard "copper bright" solution, consisting of 55% $H_3PO_4$, 25% $HNO_3$, and 20% $HC_3H_2O_3$, to etch the cavity surface for approximately one minute. We follow by rinsing with deionized water and drying with N2 gas.

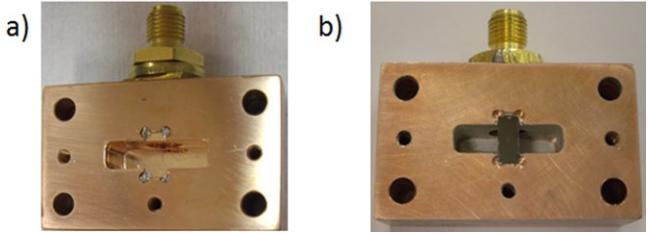

Fig. 1. (color online) (a) OFHC Cu half-cavity that has been elelctropolished. (b) Half-cavity with the walls electroplated with Sn; a transmon qubit fabricated on a Si substrate has been placed in the chip recess. The SMA connectors are threaded in for different external coupling levels between (a) and (b). The outer dimensions of the OFHC Cu block in which the cavity is machined are $3.8 \times 2.3$ cm$^2$ for the open face as pictured above with a height of 1.4 cm.

We employ an electropolishing process using a commercially available solution (*EP 3000 Chemtrec* [16]) and an in-house electropolishing setup (Fig. 2). The stainless steel electrodes, which are inert to the electropolishing solution, are cut such that they match the cavity dimensions. The cavity is held on an L-shaped anode immersed in the electropolishing solution and a cathode that is cut to fit the dimensions of the half-cavity width is inserted inside (Fig. 2). Precaution must be taken such that the cathode does not touch the cavity walls and cause a short circuit. For achieving a more uniform electric field within the cavity, the side-threaded SMA port is covered with tape and the anode must be in contact with the entire back surface of the cavity. The anode and cathode are connected to a DC power supply. During the polishing process, we have found that a manual variation of the power supply voltage between 2 V and 4 V with a period of roughly 3 minutes and a total time of about 30 minutes resulted in the smoothest surfaces that we were able to achieve (Fig. 3). For voltages below 2 V the polishing rate was too slow, while voltages beyond 4 V tended to cause the electropolishing process to become too rapid, resulting in significant oxidation of the surface and a corresponding darker color. We note that it is critical to perform the electropolishing immediately after the copper-bright etch in order to avoid any oxide formation in between the steps. After electropolishing, the cavity is solvent-rinsed once more. To quantify the improvement in surface quality, we measured the RMS roughness using a KLA-Tencor surface profilometer operated in the 2D scan mode. The average roughness of an as-machined Cu surface was 4 μm, while after electropolishing it was reduced to 0.4 μm.

In order to characterize the effects of the electropolishing on the microwave loss, we used a vector network analyzer to measure the transmission between the two SMA connectors of each cavity. A simple fit to the resonance allowed us to extract the cavity linewidth, and thus, the total quality factor Q. For an unpolished copper cavity, we obtained a room-temperature Q of about 2500. The same cavity after elelctropolishing had a quality factor of 4800. We also measured the low-temperature quality factors of the cavities by mounting them on the lower stage of a cryostat equipped with a pulse-tube cooler, which typically achieved temperatures between $2.7 - 3$ K. For the unpolished cavity we measured a Q of 8800 at 2.9 K, while after electropolishing we measured the Q at 2.9 K to be 12,000.

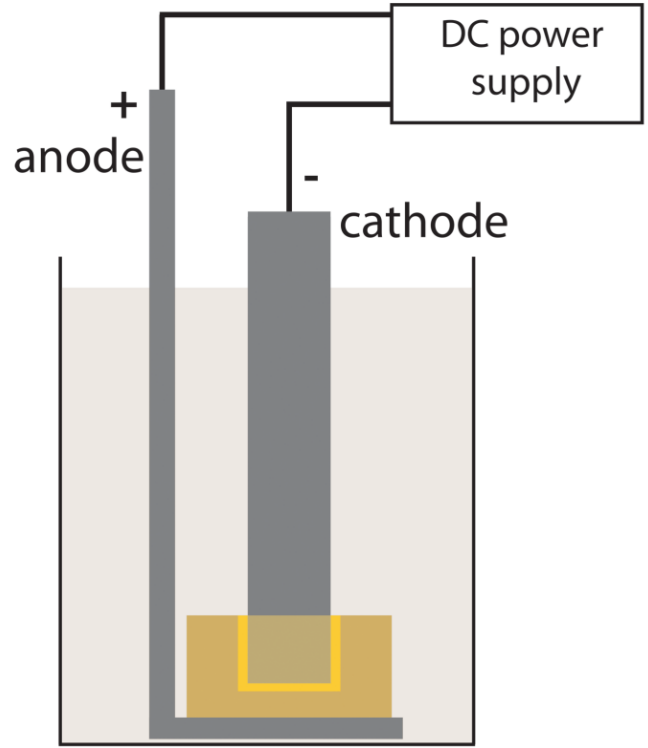

Fig. 2. (color online) Schematic design of the electropolishing setup; the cavity is shown in cross-section. As described in the text, it is essential for the anode to cover the entire back area of the half-cavity.

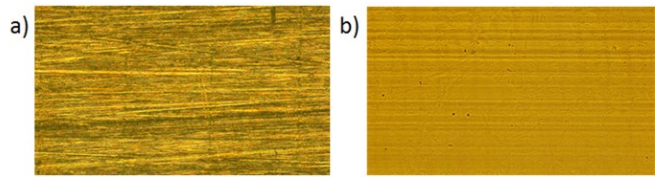

Fig. 3. (color online) Optical micrograph of an 800 X 1300 μm$^2$ area of OFHC Cu (a) as-machined; (b) after electropolishing.

To further reduce the microwave loss in the cavity walls, we chose to deposit a superconducting layer on the copper walls. Because of the narrow rectangular geometry of the cavities, traditional thin-film deposition methods, such as evaporation or sputtering, are not feasible. Thus, we chose to develop an electroplating process instead. Commercial electroplating solutions are available for depositing Sn (www.transene.com), and the superconducting transition temperature of ~3.7 K is suitable for our application. We developed an in-house setup for this electroplating process with a similar arrangement to our electropolishing system (Fig. 2), although the polarity was changed, with the cathode connected to the bottom of the half-cavity block and the anode cut to dimensions such that it fit inside the half-cavity. The electrodes were designed to provide a reasonably uniform electric field distribution inside the cavity during the plating. For the best plating results we always used fresh electroplating solution. After exploring the bias-parameter space through a series of tests, we observed the most uniform coating of Sn for bias voltages in the range between 1.5 V and 2.5 V. Higher values resulted in a more porous Sn layer with a weaker adhesion to the cavity walls. Also, a larger contact area between the cathode and the backside of the cavity block improved the coating uniformity. Nonetheless, due to geometric constraints from the cavity shape, our best Sn-coating layers to date still have some nonuniformity, with a higher thickness on the cavity walls and less at at the edges and corners. The thickness of the coated Sn is in the range of 2 to 3 μm on the cavity walls. This was determined by electroplating a test piece of Cu that was electropolished and electroplated for the same amount of time as our cavities. We used a small piece of tape to mask off part of the surface in order to produce a sharp edge for subsequent thickness measurements with a surface profilometer.

During the plating process, the Sn was also deposited on the outside surfaces of the copper cavity block, but this was easily removed afterwards with simple mechanical polishing. Following electroplating we rinse the cavity in DI water and use an ultrasonic cleaner for about one minute to remove the loose Sn whiskers that tend to form on the edges of the half-cavity. We then remove the Sn deposited in the mounting pockets for the qubit chip to allow the chip to rest directly on the copper [Fig. 1(b)] for improved thermal contact.

### III. LOW-TEMPERATURE CHARACTERIZATION OF SN-PLATED COPPER CAVITIES

At room temperature, the Sn-plated copper cavities yielded typical Q values of about 900. This is even lower than the bare copper cavities, but is not particularly surprising due to the likely level of disorder in the electroplated Sn films. We measured the low-temperature performance of Sn-plated cavities with cavity-II, initially on our pulse-tube cooler-based cryostat with a base temperature around 2.7 K. We were able to measure the temperature-dependence by turning off the pulse-tube cooler and recording the cavity transmission and temperature while the cryostat slowly warmed up. Upon warming, we observed a sharp increase in the total loss, $1/Q$, around 3.6 K, near the superconducting transition temperature for bulk Sn (Fig. 4). The significant width of the superconducting transition is likely caused by non-uniformities in the coating of the cavity surfaces. Nonetheless, the loss measured at 2.7 K corresponds to a total Q of 92,000 [15]. Also, we note that over multiple thermal cycles of Sn-plated cavities in our pulse-tube cryostat and subsequently in our dilution refrigerator, we observed no significant change in the surface quality of the Sn layers or of the cavity Q.

In order to measure the quality factor of the Sn-plated copper cavity at millikelvin temperatures, relevant for qubit experiments, we have cooled down cavity-II with no qubit present on our dilution refrigerator at Syracuse with a typical base temperature of 30 mK. On the same cooldown, we were also able to calibrate an identical set of measurement leads on the refrigerator to allow for an estimate of the baseline transmission in the absence of the cavity. From a simple fit to extract the cavity linewidth, we obtained a total Q of 124,000, which is slightly higher than the value measured at 2.7 K for this cavity, but consistent with the broad tail of the superconducting transition for these Sn-plated films. By incorporating the baseline transmission estimate and a circuit model for the cavity resonance and external circuitry [18], we were now able to separate out the external coupling and internal losses. In this way we obtained an external coupling quality factor $Q_c$ of 950,000 and an internal quality factor $Q_i$ of 143,000.

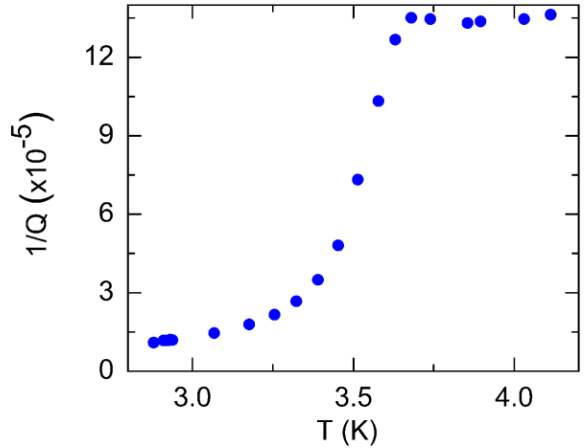

Fig. 4. (color online) Temperature dependence of total loss for an empty Sn-plated copper cavity measured across the Sn superconducting transition temperature.

Thus, the Sn-plating resulted in an increase in Q relative to the bare copper cavities by more than an order of magnitude. This reduction in the cavity linewidth is then quite helpful for achieving reasonable measurement fidelity for the dispersive readout of a qubit in such a cavity. At the same time, the quality factor we have achieved in Sn-plated copper cavities is still significantly lower than what has been measured in bulk aluminum cavities -- $Q \sim 10^6$ [8] -- or especially in aluminum cavities with an optimized geometry and surface treatment -- $Q \sim 10^9$ [19]. Besides these aluminum cavities, in the area of atomic cavity QED, even higher quality factors have been measured -- $Q \sim 10^{10}$ [20] -- with Cu cavities that have been diamond-polished then sputter-coated with superconducting Nb. Thus, it should be possible to start with a copper cavity base for coupling to transmon qubits and reach even higher quality factors than what we have achieved so far.

## IV. QUBIT MEASUREMENTS

In addition to characterizing the low-temperature quality factor of empty Sn-plated copper cavities, we have also performed a series of experiments with transmon qubits in two of these cavities at millikelvin temperatures. The qubits have a similar geometry to those studied in Ref. [12], with Al capacitor paddles that are $350 \times 700$ $\mu m^2$ each with a 50 $\mu m$ separation between the paddles and a single Al/AlOx/Al Josephson junction. Qubit A was fabricated on high-resistivity Si with $\rho > 10$ k$\Omega$-cm and a chip size of $3.2 \times 6.8$ mm$^2$. Qubit B was fabricated on sapphire with a chip size of $2.0 \times 6.8$ mm$^2$. Although both qubits had the same geometry, Qubit A had an array of 20 x 20 $\mu m^2$ holes with a spacing of 25 $\mu m$ on each capacitor paddle; Qubit B had only one row of 20 x 20 $\mu m^2$ holes along the inner edge of each paddle with a spacing of 60 $\mu m$. In both cases, the holes were intended to avoid the trapping of Abrikosov vortices, which can lead to excess loss in superconducting microwave resonant circuits [21], although the scheme for the hole configurations was not a focus of our present study of Sn-plated copper cavities.

We measured these qubits and cavities on dilution refrigerators at both Syracuse and IBM using a conventional configuration of cold microwave attenuators, isolators, and filters for thermalizing the measurement cables. We employed standard cQED techniques [14, 22] for measuring the qubit state by probing the microwave transmission through the cavity. In addition, the devices were shielded from stray infrared radiation by covering the cavity block with eccosorb® strips [23, 24], copper tape, and aluminized mylar foil, as well as an eccosorb-lined cryogenic magnetic shield mounted to the cold-plate of the refrigerator.

In our initial experiments, we cooled down Qubit A in cavity-I with Sn-plating. The presence of the Si chip in the center of the cavity lowered the resonance frequency for the TE101 mode to 10.917 GHz at 30 mK. The SMA connectors were inserted to a depth that resulted in a low-temperature total quality factor of 42,600. Using a conventional cQED spectroscopy technique, applying a 40 $\mu s$ microwave signal while incrementing through its frequency followed by a second pulse near the cavity resonance frequency to probe the cavity transmission, we identified the qubit transition frequency between the ground and first-excited states to be 4.3362 GHz. We then measured the relaxation time $T_1$ by monitoring the decay of the qubit population following the application of a $\pi$-pulse and we obtained $T_1 = 33$ $\mu s$ [Fig. 5(a)]. We used a standard pulse sequence to perform a Ramsey fringe experiment using two detuned $\pi/2$ pulses with a variable delay time for extracting a coherence time $T_2^*$ of 41 $\mu s$ [Fig. 5(b)]. These measurements were performed in the high-power, many-photon regime [14] at IBM. On a previous cooldown at Syracuse several weeks earlier, we measured the same qubit and cavity in the low-power, dispersive regime and obtained similar values: $T_1 = 36$ $\mu s$ and $T_2^* = 42$ $\mu s$ (not shown). This suggests that the measurement setups in both locations provide comparable levels of shielding and filtering and that the cavity and chip-mounting scheme is robust. Furthermore, the technique for measuring the cavity transmission – the low-power, dispersive approach or the high-power, many-photon regime – do not appear to influence the qubit coherence significantly.

Qubit B was mounted in cavity-II with Sn-plating and cooled down on the Syracuse refrigerator. With the sapphire chip present, the bare cavity resonance was lowered to 12.274 GHz. With a weaker coupling between the cavity and the external circuitry compared to the previous cooldown of Qubit A in cavity-I, we measured a total Q of 89,000. Because we did not have an independent calibration of the transmission through the measurement line, we were unable to separate out the external and internal losses for this cooldown. With standard cQED spectroscopy, we observed the ground-to-first excited state transition at 5.9903 GHz. Following the same pulsed-measurement schemes as with Qubit A, we measured cavity transmission in the low-power, dispersive regime and obtained $T_1=31$ $\mu s$ and $T_2^*=39$ $\mu s$ (Fig. 6). Experimental constraints prevented us from running a Ramsey fringe measurement out to longer pulse separations for this particular qubit. Thus, the uncertainty in the $T_2^*$ value from the fit is quite significant, +/-7 $\mu s$. Nonetheless, the decay exhibited in Fig. 6(b) is rather slow, indicating a qubit with good coherence properties, even if this particular fit prevents us from extracting a more precise value of $T_2^*$.

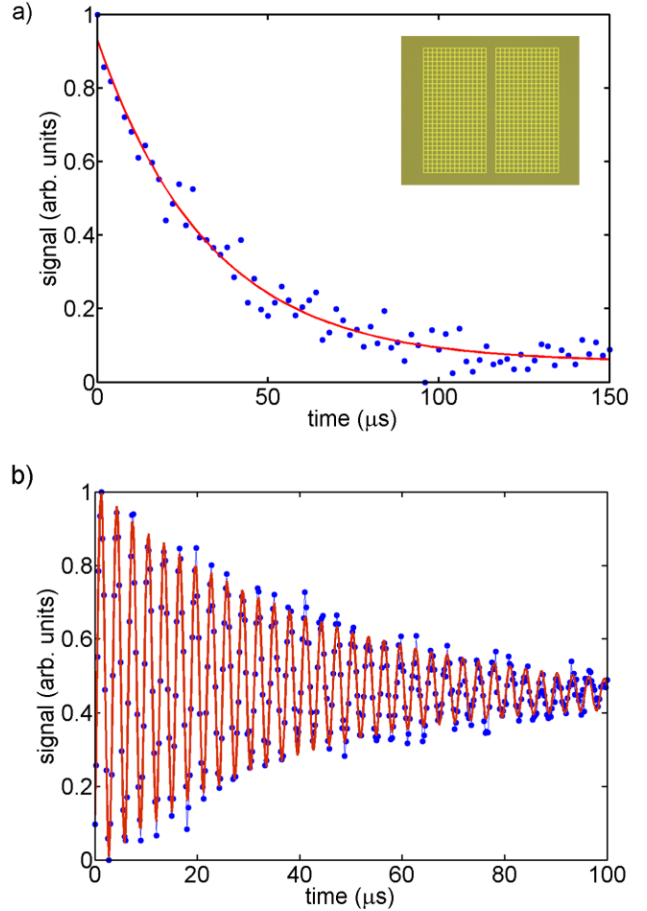

Fig. 5. (color online) (a) Energy relaxation time $T_1$ of Qubit A (Si substrate) mounted in a Sn-plated copper cavity. Fit to exponential decay corresponds to $T_1 = 33$ $\mu s$. The inset shows an optical micrograph of the qubit. (b) Coherence time measurement (Ramsey fringes) for the same qubit/cavity corresponding to $T_2^*$ of 41 $\mu s$ (based on Fig. 3 in Ref. [15]).

Based on our measurements of the two different qubits in multiple cooldowns, our coherence times in the two different Sn-plated copper cavities are promising and repeatable. Also, it is important to note that similar qubit chips from the same sapphire wafer that produced Qubit B were characterized at IBM in aluminum cavities with a nearly identical geometry to our Sn-plated copper cavities. The measured qubit coherence times in the aluminum cavities were also in the 30 μs range. Thus, the use of copper over aluminum for the cavity structure does not appear to be a limiting factor in determining the qubit coherence for these particular devices. Nonetheless, these coherence times are not yet up to the record times in excess of 100 μs that have been reported for transmon qubits in three-dimensional waveguide cavities [25, 26]. Both of these efforts involved aluminum cavities, which would have had internal quality factors much larger than those of our Sn-plated copper cavities.

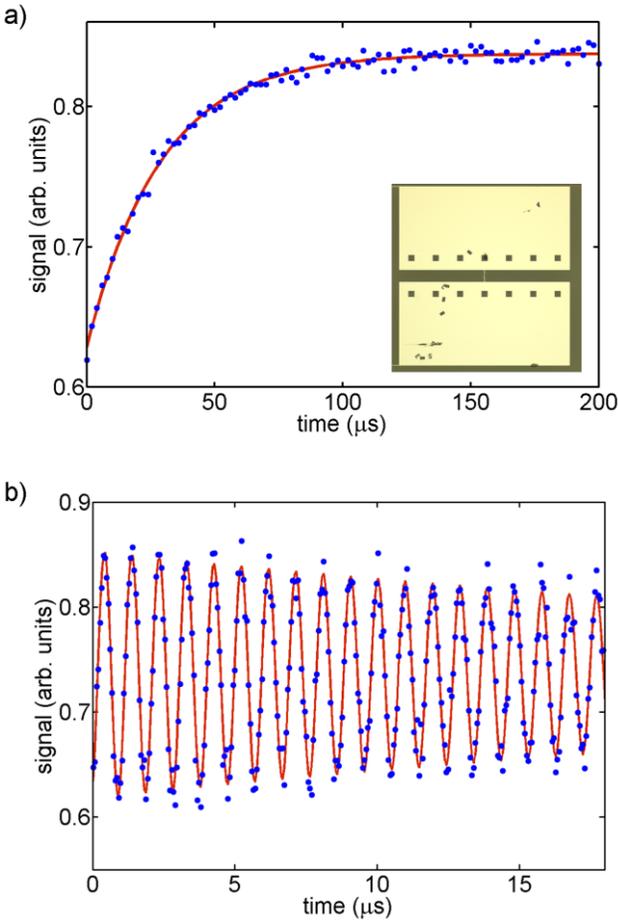

Fig 6. (color online) (a) Energy relaxation measurement of Qubit B (sapphire substrate) mounted in a Sn-plated copper cavity. The inset shows an optical micrograph of the measured qubit. (b) Ramsey fringe measurement for the same qubit/cavity.

However, for a sufficient detuning between the qubit and cavity, even a bare copper cavity with a much lower Q can allow qubit coherence times approaching 100 μs [12]. Thus, after employing sufficiently large cavity-qubit detuning and following the state of the art for thermalization of the measurement leads and shielding of the qubit and cavity, it appears that cavity loss alone does not determine qubit performance for this system. Instead, there are likely multiple remaining factors that limit the qubit coherence times and some of these may not even be identified yet. One likely candidate source includes dissipation on the qubit chip itself. Although the approach to placing the transmon in a three-dimensional hollow cavity eliminates much of the substrate that is typical for experiments with planar structures and coplanar-waveguide resonators, there is still a small chip on which the transmon is fabricated. Improvements in the materials processing and patterning of the superconducting films may address this to some extent [27, 28], but there may be other details of the chip-processing that could play a role. Besides loss on the chip itself, the nature of the interface between the chip and the recess that it mounts into in the cavity structure may also limit the ultimate qubit coherence. If the chip is not anchored sufficiently well to the cavity structure, the thermal contact resistance at this interface might limit the efficiency with which heat in the qubit chip can escape to the cavity block and get absorbed by the refrigerator.

For our measurements presented here on Sn-plated copper cavities, we have used the same techniques for mounting the qubit chips in the cavity recesses that we and others have used for measurements of transmon qubits in aluminum or copper cavities [12]. Before mounting the qubit chip in the cavity recess, we clean the surface with fine-grade sandpaper followed by a solvent cleaning. After pressing the chip into the recess, we use small amounts of indium placed in the enlarged corners of the recess to anchor the chip in place (Fig. 1).

With this approach to mounting the microfabricated chip in a macrofabricated three-dimensional cavity that we and others have employed [12], it is clearly quite difficult to characterize the nature of the thermal contact resistance between the chip and the cavity and it is rather challenging to achieve a consistent and robust approach for placing the qubit chips in the cavities. In fact, thermal contact resistances at low temperatures between certain dissimilar materials have been studied extensively [29], but we are not aware of any previous work on the materials and geometry in the typical 3D cavity/qubit system. Nonetheless, because of the similar mounting techniques, we expect that this thermal contact resistance is not substantially different for qubits mounted in our Sn-plated copper cavities compared to aluminum cavities. This is certainly an important area to explore in the future, perhaps with dedicated experiments to measure thermal time constants with new schemes for applying varying amounts of pressure between the cavity and qubit chip. Ultimately, once thermal interface properties between qubit chips and cavities are better understood, having a copper base for the cavity structure should be helpful in this direction, as one would be starting with a well-thermalized contact surface between the cavity and qubit chip.

While our measurements presented here demonstrate that qubits in our Sn-plated copper cavities exhibit comparable coherence times to similar devices measured in aluminum cavities, we have not yet performed an extensive investigation of qubit thermalization through measurements of the qubit excited-state population. During our spectroscopic measurements we have scanned across the frequency range where we would expect the transition from the first to second-excited state of the qubit to occur, based on our measured qubit anharmonicity. We have not observed any spectral feature above the noise floor of our spectroscopy, thus confirming that our efforts to thermalize our leads and shield the qubit from stray IR radiation are effective for avoiding an excessive thermal population of the qubit excited state [23]. In future experiments to probe qubit thermalization for different cavity materials and qubit-cavity mounting techniques, we could employ a more sophisticated technique to quantify the effective qubit temperature [30].

## V. Conclusion

We have demonstrated that the surface treatment of copper waveguide cavities through electropolishing and electroplating with superconducting tin can lead to significant reductions in the low-temperature microwave surface loss when compared to bare copper cavities. Furthermore, the plated tin films are robust over multiple cooldowns. Based on related work in the field of atomic cavity-QED [20], we expect that further improvements in the quality factor of these copper cavities should be possible with optimized cavity geometries, materials, and coating techniques. We have also performed measurements with two different superconducting transmon qubits in these polished and plated cavities and obtained good coherence times. Although the qubit performance so far has not been quite up to the current state of the art for such qubits in waveguide cavities, our measurements in Sn-plated copper cavities have yielded comparable coherence performance to devices from the same fabrication run measured in aluminum cavities. Currently there are a variety of possible mechanisms that could limit the qubit performance, including losses on the qubit chip itself and the quality of the interface between the chip and its mounting points in the cavity structure. It is possible that improved methods for anchoring the qubit chip to the cavity combined with the good thermal properties of the copper cavity base from structures similar to those presented here may provide a useful scheme for removing heat from the qubit chip and extending coherence times.

Such thermalization issues will likely be even more of a concern in future implementations of larger systems with multiple qubits and three-dimensional cavities. Recent experiments have already been performed on two transmon qubits in a waveguide cavity [31] and two three-dimensional cavities coupled to one qubit [32]. Although both of these experiments utilized aluminum cavities, significantly larger structures with many qubits and cavities [33], for example, for implementing a surface code architecture [34, 35], will have substantial thermal demands for cooling the cavity walls and qubit chips inside. Thus, copper cavities with coated superconducting walls will remain an attractive system.

## Acknowledgment

We acknowledge useful discussions and suggestions from H. Paik, S. Poletto, C. Rigetti. We also thank S. Sorokanich for help with the initial development of the electropolishing process. This research was funded by the office of the Director of National Intelligence (ODNI), Intelligence Advanced Research Projects Activity (IARPA), through the Army Research Office under grant No. W911NF-10-1-0324. All statements of fact, opinion or conclusions contained herein are those of the authors and should not be construed as representing the official views or policies of IARPA, the ODNI, or the U.S. Government.